\documentclass[aps,prf,floats,amsmath,amssymb,longbibliography,showpacs,floatfix]{revtex4-1}
\usepackage[dvipsnames,rgb,dvips]{xcolor}
\usepackage{graphicx,overpic}
\usepackage{siunitx}
\usepackage{bbold}
\usepackage{mathrsfs}

\newcommand{\ve}[1]{\ensuremath{\mbox{\boldmath$#1$}}}
\newcommand{\ma}[1]{\ensuremath{\mathbb{#1}}}

\begin{document}
\title{Critical charges for droplet collisions}
\author{A. Dubey$^1$, G. P. Bewley$^2$, K. Gustavsson$^{1}$, B. Mehlig$^{1}$}
\affiliation{
\mbox{}$^1$Department of Physics, Gothenburg University, 41296 Gothenburg, Sweden\\
\mbox{}$^2$Sibley School of Mechanical and Aerospace Engineering, Cornell University, USA}

\begin{abstract}
The collision efficiency of uncharged micron-sized water droplets in air is determined by the breakdown of hydrodynamics at droplet
separations of the order of the mean-free path, by van-der-Waals forces, or a combination of the two. In contrast, electrostatic forces determine the collision efficiency of charged droplets if the charge is large enough. To find the charge for which the transition to charge-dominated collisions occurs, we computed the collision efficiency of charged, hydrodynamically-interacting droplets settling in quiescent air,  including the breakdown of hydrodynamics at small interfacial distances.  For oppositely charged droplets, the transition occurs when a saddle point of the relative droplet-dynamics exits the region where the hydrodynamics breaks down.
For droplets with radii $16\,\mu$m and $20\,\mu$m, this occurs at $\sim 10^3$ elementary charges $e$. For smaller charges, the collision efficiency depends upon the Kn number (defined as the ratio of the mean-free-path of air to the mean droplet radius), whereas for larger charges it does not.
For  droplets charged with the same polarity, the critical charge is $\sim 10^4\,e$ for the above radii. 
\end{abstract}
\maketitle

\section{Introduction}
How do  micron-sized cloud droplets grow? This question relates to the fundamental mechanisms that determine droplet-size
distributions in atmospheric clouds \cite{pruppacher1997microphysics,shaw2003particle,devenish2012droplet}. In clouds with droplets of different sizes, droplet collisions
occur due to differential settling. This mechanism can cause rapid droplet growth \cite{berry1974analysis,grabowski2013growth}. An open question is how this process is initiated, how droplet-size differences develop in the first place. Saffman \& Turner explained that micron-sized droplets of similar sizes can collide  if turbulent strains bring them together \cite{saffman1956collision}, but this process is very slow on average. Since the sequence of droplet collisions in a cloud is  random, the collision times are essentially Poisson distributed, and fluctuations may nevertheless result in some droplets that grow very rapidly~\cite{kos2005fluctuations}. 

An uncertainty in these considerations is the collision efficiency, the ratio of the collision cross section of hydrodynamically interacting droplets to the geometric collision cross section $\propto (a_1+a_2)^2$, for two droplets with radii $a_1$ and $a_2$. The point is that hydrodynamic interactions cause the paths of approaching droplets
to curve around each other~\cite{kim2005microhydrodynamics}, reducing the cross section.  This hydrodynamic repulsion diverges at very small distances, apparently preventing the approaching droplets from colliding at all. The  divergence is unphysical though, because the hydrodynamic approximation breaks down at droplet separations of the order
of the mean free path of air \cite{sundararajakumar1996noncontinuum}. As a result of this non-continuum effect, the collision efficiency depends on the Knudsen number Kn, equal to the mean free path divided by the droplet radius. 

Hydrodynamic interactions and the breakdown of hydrodynamics result in a significant reduction of the Saffman-Turner collision rate \cite{dhanasekaran2021collision}. Its dependence on Kn and on the non-dimensional relative settling speed can be understood in terms of an intricate sequence of bifurcations
of the collision dynamics \cite{dubey2022bifurcations}. Many other factors are thought to play a role for the collision dynamics of settling droplets. Refs.~\cite{dhanasekaran2021collision,dubey2022bifurcations} neglected particle inertia. This works for very small droplets, but particle inertia allows larger droplets to detach from the flow, leading to larger collision rates \cite{sundaram1997collision,falkovich2002acceleration,gustavsson2016statistical}. Klett \& Davis \cite{klett1973theoretical} showed that the collision rate
of similar-sized droplets depends sensitively on the particle Reynolds number, a measure of the effect of convective fluid inertia upon the disturbance flow caused by the settling droplets. 
The reason is that the Stokes problem for the collision dynamics is degenerate: equal-sized droplets neither approach nor diverge. Convective
fluid inertia breaks this degeneracy \cite{candelier2007relative}. Van-der-Waals forces are short-range attractive forces due to polarisation of the  water molecules comprising the droplets \cite{rother2022gravitational} that can increase the collision rate at small Knudsen numbers. 

Small water droplets tend to carry excess electrical charges, sometimes positive, sometimes negative. It is clear that electrostatic forces either reduce or increase the collision efficiency, depending on whether the two droplets have excess charges of the same polarity or not. But how much charge is required to make a significant difference to the collision dynamics? This question is motivated in part by the fact that water droplets in atmospheric clouds are charged \cite{harrison2015microphysical}. The question has a long history, but it has not been resolved. Davis \cite{davis1965theoretical} found
for instance that  more than 800 elementary charges $e$ are required for the Coulomb force to change the collision efficiency for  radii of the order of $20\,\mu$m,  
 the typical size of cloud droplets, yet \citet{tinsley2014comments} suggest that it may take at least $10^4\,e$.  These differences are not surprising,
 given that the estimates are based on highly idealised models 
  that either do not account for how the hydrodynamic forces change as the droplets come close \cite{tinsley2014comments,paluch1970theoretical,schlamp1976numerical,schlamp1979numerical}, or do not consider how the hydrodynamic approximation breaks down below the mean free path \cite{davis1965theoretical,tinsley2014comments,paluch1970theoretical,schlamp1976numerical,schlamp1979numerical}, causing the predictions to fail \cite{abbott1974experimental}.  
  
Magnusson {\em et al.} \cite{magnusson2021collisions} used models for hydrodynamic and electrical interactions valid at large droplet separations to analyse collisions of strongly charged $20\mu$m-droplets, with opposite polarities, settling in quiescent air. In this regime, the collision efficiency is determined by a stable manifold of a saddle point of the relative droplet dynamics. At the saddle point, the larger droplet travels below the smaller one -- so that Coulomb attraction, hydrodynamic interactions, and the difference of the gravity forces cancel, resulting in a steady state. When the charges are strong enough, the droplets are far apart at this equilibrium point. In this case, non-continuum effects do not matter, and it is sufficient to consider the large-distance asymptotes of the hydrodynamic interactions.  These approximations break down at smaller charges. Therefore we cannot use the model analysed by  Magnusson {\em et al.} \cite{magnusson2021collisions}  to estimate the smallest amount of charge  needed to significantly change the collision dynamics of micron-sized droplets.

In order to determine the collision efficiencies of weakly charged droplets settling under gravity, we performed numerical simulations of the collision dynamics, taking into account particle inertia, hydrodynamic interactions not only at large but also
at small droplet separations, as well as their regularisation by non-continuum effects due to the breakdown of hydrodynamics at interfacial distances smaller than the mean free path. We modeled the droplets as perfect conductors \cite{lekner2012electrostatics}, which allowed us to consider the effect of induced charges that affect the electrostatic force at small separations. This is important, because induced charges can cause droplets to attract each other even though they are charged with the same polarity \cite{khain2004rain,fletcher2013effect}. We found that the collision efficiency exhibits two qualitatively different regimes: a small-charge regime where non-continuum effects determine collisions and a large-charge regime where electrical forces determine collisions. We determined how much charge is needed to affect this qualitative change in the collision dynamics. 
The transition region is quite broad as the charge varies, but the qualitative change occurs when a saddle point of the relative droplet-dynamics exits the spatial region where non-continuum effects dominate. 
The critical charge depends on whether the droplets are charged with the same or with opposite polarities.  
For two droplets with radii $16\,\mu$m and $20\,\mu$m charged with the opposite polarities, the transition occurs at $\sim 10^3$ elementary charges $e$.
When the droplets have excess charges with the same polarity, $\sim 10^4\,e$ are required. 
In addition, a bifurcation occurs at about $1.5\times 10^4$ elementary charges, leading to two saddle points above the collision sphere.
Droplets approaching near the symmetry axis -- the $R_3$-axis in Fig.~1({\bf a}) --  spend a long time in the vicinity of  one of the saddle points, before they diverge along its unstable manifold. 

 \section{Model}
 \label{sec:model}
\subsection{Equations of motion}

\begin{figure}[t]
\begin{overpic}[scale=.4]{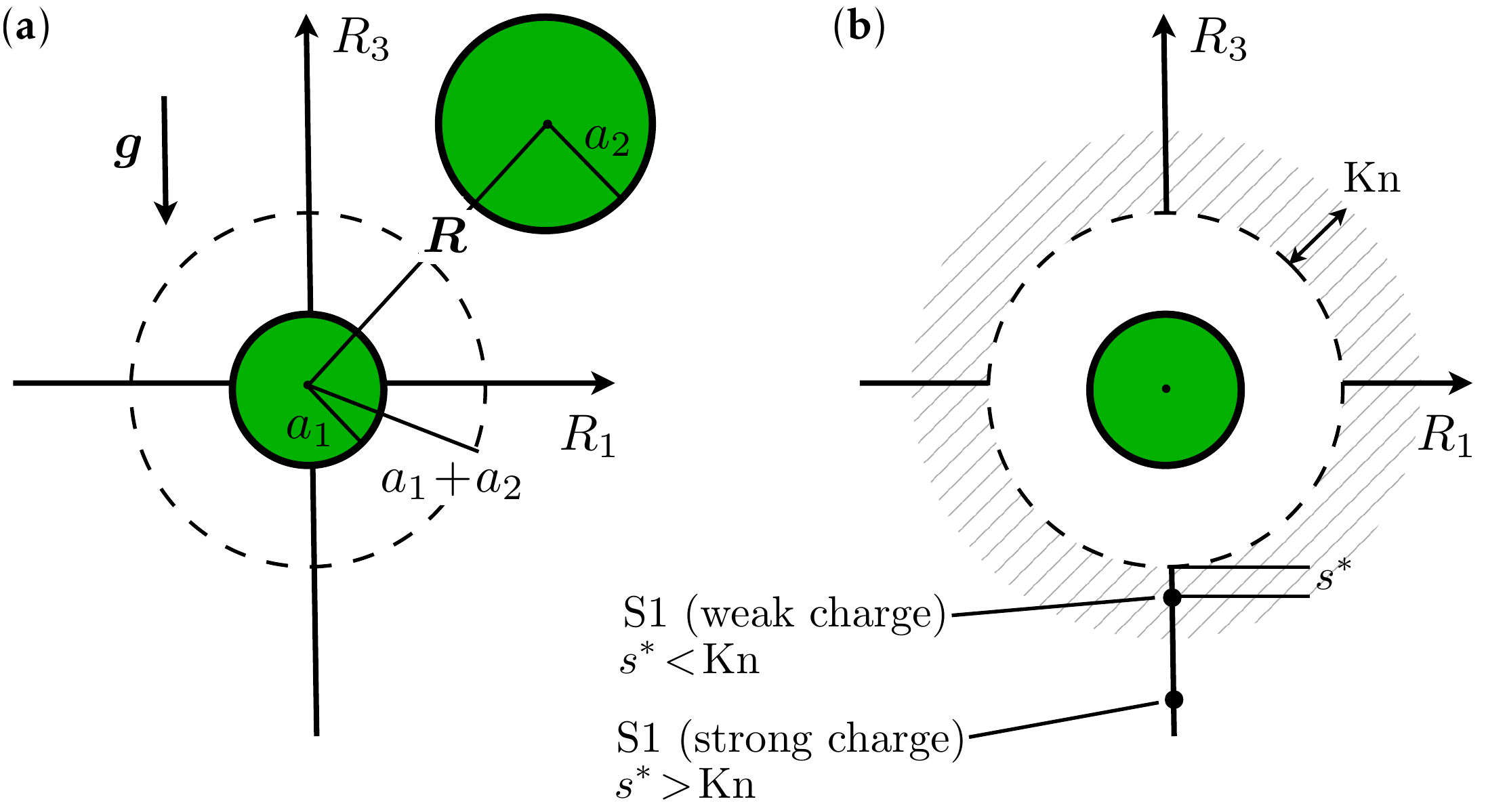}
\end{overpic}
\caption{\label{fig:schematic}  Panel ({\bf a}) shows a schematic of two droplets with radii $a_1<a_2$ settle in a quiescent fluid. The separation vector between their centres-of-mass is $\ve R$. Also shown is the collision sphere around the smaller droplet (dashed). The droplets collide if the centre-of-mass of the larger one hits the collision sphere. 
Gravity points in the negative $R_3$-direction. Panel ({\bf b}) illustrates two regimes of the collision dynamics, distinguished by the location
of the fixed point  S$1$ (see text).
For weak charges,  S$1$ is inside the region where hydrodynamics breaks down (hashed area), at non-dimensional
distance $s^\ast\equiv R^\ast-2<{\rm Kn}$. For strong charges, the fixed point lies outside, $s^\ast>{\rm Kn}$.}
\end{figure}
The motions of two spherical droplets $i=1,2$ are determined by Newton's equations,
\begin{subequations}
\begin{align}
 \dot{\ve x}^{(i)}&= \ve v^{(i)}\,, \label{eq:model_a}\\
 \label{eq:vdot}
 \dot{\ve v}^{(i)} &=[\ve F_{\rm g}^{(i)}+\ve F_{\rm h}^{(i)} + \ve F_{\rm e}^{(i)}]/m_i \,,\\ 
 \dot{\ve \omega}^{(i)} &= \ve T_{\rm h}^{(i)}/I_i\,. \label{eq:model_c}
\end{align}
\end{subequations}
Here $\ve x^{(i)}$ and $\ve v^{(i)}$ are the position and velocity of droplet $i$, and ${\ve \omega}^{(i)}$ is its angular velocity. 
Droplet $i$ has mass $m_i=\tfrac{4\pi}{3}\varrho_{\rm w} a_i^3$ and moment of inertia $I_i={\tfrac{2}{5}m_ia_i^2}$, where $a_i$ is the radius of droplet $i$, and $\varrho_{\rm w}$ 
is the mass density of water. On the r.h.s. of Eq.~(\ref{eq:vdot}), $\ve F_{\rm g}^{(i)} = m_i \ve g$ is the gravity force with gravitational acceleration $\ve g$, and
 $\ve F_{\rm e}^{(i)}$ is the electrostatic force on droplet $i$, $\ve F_{\rm h}^{(i)}$ is the hydrodynamic force, and $\ve T_{\rm h}^{(i)}$ is the hydrodynamic torque.
 The droplets settle through quiescent air. We denote their centre-of-mass separation by $\ve R= \ve x^{(2)}-\ve x^{(1)}$  (Fig.~\ref{fig:schematic}),
 and their relative velocity by $\ve V=  \ve v^{(2)}-\ve v^{(1)}$. 
 
 We non-dimensionalise the problem as follows. As a length scale we use $\overline{a}=(a_1+a_2)/2$.
 Velocities are non-dimensionalised with the differential settling speed of non-interacting droplets,  $V_0= g {|}\gamma_2^{-1}-\gamma_1^{-1}{|}$.
 Here $\gamma_i = (9 /2) ( \varrho_{\rm f} / \varrho_{\rm w}) \nu/a_i^2$ is the Stokes  constant of droplet $i$, with fluid-mass density $\varrho_{\rm f}$
 and kinematic viscosity $\nu$. Time is non-dimensionalised with the characteristic time scale $\tau_{\rm c}=\overline{a}/V_0$.
  The non-dimensional parameters are the Stokes number  ${\rm St} = (\widetilde\gamma \tau_{\rm c})^{-1}$ (particle inertia), and the Coulomb number
  ${\rm Cu}=2 k_{\rm e} |q_1 q_2|/[\widetilde{m} (a_1+a_2) V_0^2]$ (electrical charge). Here $\widetilde \gamma= (9 /2) ( \varrho_{\rm f} / \varrho_{\rm w}) \nu/\overline{a}^2$ and
 $ \widetilde m=(4/3)\pi \overline{a}^3 \varrho_{\rm w}$ are Stokes constant and droplet mass based on the mean droplet radius $\overline{a}$. The Coulomb number
 is the  ratio of the  Coulomb energy upon contact and $\widetilde{m}V_0^2/2$. 
 Related parameters (ratio of Coulomb to kinetic energy) were used by \citet{hidy1965remarks} to  quantify the effect of electric charge on Brownian coagulation, and in Refs.~\cite{lu2010clustering_a,lu2010clustering_b,lu2015charged,magnusson2021collisions} to parameterise the importance of charge on the dynamics of micron-sized water droplets in turbulent air.
 The remaining three non-dimensional
 parameters are the Knudsen number  ${\rm Kn}=\ell/\overline{a}$, the ratio of the
 mean free path of air $\ell$ to the mean droplet radius, the radius ratio $\lambda = a_1/a_2$, and the charge ratio $\lambda_q = q_1/q_2$. 
 The charge ratio can take positive or negative values, depending on the polarities of the excess charge on the droplets.

 \subsection{Electrostatic force}\label{sec:electrostatic} We assume that the droplets are good conductors, so that we may use Lekner's results for the  electrostatic force between two charged conducting spheres \cite{lekner2012electrostatics}.  We outline the main steps of
 the derivation in Ref.~\cite{lekner2012electrostatics} here, using dimensional units. 
The electrical potential energy $\mathscr{W}$ of two conductors with charges $q_i$ and potentials $V_i$ is given by
\begin{align} \label{eq:electricalW}
 \mathscr{W}= \frac{ k_{\rm e}}{2} \,(q_1 V_1 +q_2V_2)\,,
\end{align}
where $k_{\rm e}$ is the Coulomb constant.
The charges $q_i$ are linearly  related to the electrical potentials~$V_j$,
\begin{align}\label{eq:electricalC}
 \begin{bmatrix}
  q_1 \\
  q_2
 \end{bmatrix} = \begin{bmatrix}
 C_{11} & C_{12} \\
 C_{21} & C_{22}
 \end{bmatrix}\begin{bmatrix}
  V_1 \\
  V_2
 \end{bmatrix}\,,
\end{align}
with capacitance coefficients $C_{ij}$ \cite{jackson1975classical}.
The capacitance coefficients  depend on the radii of the spheres, $a_1$ and $a_2$, and upon the centre-of-mass distance between them as follows:
\begin{subequations}
\label{eq:lekner}
\begin{align}
 C_{11} &= a_1 a_2 \sinh (U) \sum_{n=0}^\infty [a_1 \sinh (nU) + a_2 \sinh((n+1)U)]^{-1}\,,\\
 C_{22} &= a_1 a_2 \sinh (U) \sum_{n=0}^\infty [a_2 \sinh (nU) + a_1 \sinh((n+1)U)]^{-1}\,,\\
 C_{12} &=C_{21}= - a_1 a_2 \frac{\sinh (U)}{R} \sum_{n=1}^\infty [\sinh (nU)]^{-1}\,,
 \end{align}
 \end{subequations}
with $ \cosh (U) = (R^2-a_1^2-a_2^2)/(2a_1a_2)$.
Eliminating the potentials in Eq.~(\ref{eq:electricalW}) by means of (\ref{eq:electricalC}), one obtains the electrostatic potential energy $\mathscr{W}$, and by differentiation of $\mathscr{W}$
w.r.t. $R$ one obtains the electrostatic force. Truncating the sums in Eqs.~\eqref{eq:lekner} allows us to compute the electrical force when droplets are far apart. We denote this asymptote
by  $F_{{\rm e},\rm far}$.
When $R\gg a_1+a_2$, the sums in Eq. (\ref{eq:lekner}) converge rapidly. When the centre-of-mass distance is close to $a_1+a_2$, however, this is not the case. 
Lekner derived an approximate expression for the force valid at small distances $s=R-a_1-a_2$, Eq.~(3.4) in Ref.~\cite{lekner2012electrostatics}:

\begin{equation}
\label{eq:lekner2}
F_{{\rm e},{\rm near}}= - k_e\,\frac{a_1+a_2}{a_1a_2 s}\frac{ \left[ \left(\psi(\tfrac{1}{1+\lambda})+\gamma\right) q_2 -\left(\psi(\tfrac{\lambda}{1+\lambda})+\gamma\right)q_1 \right]^2}{\left[\left(\psi(\tfrac{1}{1+\lambda})+\gamma\right) \left\lbrace2\gamma + \log[\tfrac{2a_1a_2}{(a_1+a_2)s}]\right\rbrace+\left(\psi(\tfrac{\lambda}{1+\lambda})+\gamma\right) \left\lbrace-2 \psi(\tfrac{1}{1+\lambda}) + \log[\tfrac{2a_1a_2}{(a_1+a_2)s}]\right\rbrace\right]^2}.
\end{equation}
valid as $s\to 0$.  Here  $\psi(x)$ is the digamma function, and $\gamma=0.5772\ldots $ is Euler's  constant.  In order to obtain an approximation valid uniformly in $R$, the two asymptotes  are matched at the interfacial distance $s_{\rm m}$ where the absolute difference between the two asymptotes reaches a minimum, 
using exponential smoothing of the form
\begin{equation}
 F_{{\rm e},\rm uv}= F_{{\rm e},\rm near}\, {\rm e}^{-s/{s_{\rm m}}}+ F_{{\rm e},\rm far}\, \left(1- {\rm e}^{-s/{s_{\rm m}}}\right)\,.
\end{equation}

 \subsection{Hydrodynamic forces}
Since the droplets settle in still air, there is no undisturbed fluid velocity or vorticity. The hydrodynamic forces $\ve F_{\rm h}^{(i)}$ and torques $\ve T_{\rm h}^{(i)}$ are calculated in the steady Stokes approximation as linear functions of the translational and angular velocities  \cite{kim2005microhydrodynamics},
\begin{align} \label{eq:resistance}
\begin{bmatrix}
\ve F_{\rm h}^{(1)}\\
\ve F_{\rm h}^{(2)}\\
\ve T_{\rm h}^{(1)}\\
\ve T_{\rm h}^{(2)}
\end{bmatrix} = \begin{bmatrix}
\ma A_{11} & \ma A_{12} &  \widetilde{\ma B}_{11} & \widetilde{\ma B}_{12}\\
\ma A_{21} & \ma A_{22} &  \widetilde{\ma B}_{21} & \widetilde{\ma B}_{22}\\
\ma B_{11} & \ma B_{12} &  \ma C_{11} & \ma C_{12}\\
\ma B_{21} & \ma B_{22} &  \ma C_{21} & \ma C_{22}
\end{bmatrix} \begin{bmatrix}
\ve v^{(1)}\\
\ve v^{(2)}\\
\ve \omega^{(1)}\\
\ve \omega^{(2)}
\end{bmatrix}.
\end{align}
 The elements of the resistance tensors $\ma A, \ma B,$ and $\ma C$ are not independent. They are related by symmetry.  Some of the symmetry relations 
 follow from the reciprocal theorem, while others are due to the spherical shape of the droplets \cite{jeffrey1984calculation}. A consequence of the symmetries is that the elements. 
 of each resistance tensor can be written in terms of at most two scalar functions: the radial resistance function $X_{{\alpha\beta}}$ and the tangential resistance function $Y_{{\alpha\beta}}$. In the notation of Refs.~\cite{jeffrey1984calculation,kim2005microhydrodynamics}
 the resistance tensors read
 \begin{subequations}
\label{eq:resistance_functions}
 \begin{align}
 A^{\alpha \beta}_{ij} &= -6 \pi \mu \frac{a_\alpha+a_\beta}{2} [X^A_{\alpha \beta} \, \widehat R_i \widehat  R_j + Y^A_{\alpha \beta} (\delta_{ij}-\widehat R_i\widehat R_j)], \\
 B^{\alpha \beta}_{ij} &= -4 \pi \mu \left(\frac{a_\alpha+a_\beta}{2}\right)^2 [ Y^B_{\alpha \beta} \epsilon_{ijk} \widehat R_k], \\
 \widetilde B^{\alpha \beta}_{ij} &= B^{\beta \alpha}_{ji}, \\
  C^{\alpha \beta}_{ij} &= -8 \pi \mu \left(\frac{a_\alpha+a_\beta}{2}\right)^3 [X^C_{\alpha \beta} \, \widehat R_i \widehat R_j + Y^C_{\alpha \beta} (\delta_{ij}-\widehat R_i\widehat R_j)].
\end{align}
\end{subequations}
The non-dimensional  resistance functions $X_{{\alpha \beta}}$ and $Y_{{\alpha \beta}}$ depend upon the non-dimensional centre-of-mass distance $R$ between the droplets, non-dimensionalised by 
the mean droplet radius $\overline{a}=(a_1+a_2)/2$, and upon the radius ratio $\lambda = a_1/a_2$. Moreover, $\widehat R_j \equiv R_j/R$. 
In total there are 20 resistance functions in Eqs.~\eqref{eq:resistance_functions}: $X^A_{\alpha \beta}, Y^A_{\alpha \beta}, Y^B_{\alpha \beta}, X^C_{\alpha \beta}$, and $Y^C_{\alpha \beta}$, all with indices $\alpha, \beta = 1,2$. Ten out of these are given in the main text of Ref.~\cite{jeffrey1984calculation}. The remaining ten functions are obtained using the relations \cite{jeffrey1984calculation}
\begin{subequations}
\label{eq:restensors}
\begin{align}
 X^A_{\alpha \beta} (R,\lambda) &= X^A_{\beta \alpha }(R,\lambda) = X^A_{(3-\alpha )(3-\beta) }(R,1/\lambda)\,, \\
 Y^A_{\alpha \beta} (R,\lambda) &= Y^A_{\beta \alpha }(R,\lambda) = Y^A_{(3-\alpha )(3-\beta) }(R,1/\lambda)\,, \\
 Y^B_{\alpha \beta} (R,\lambda) &= - Y^B_{(3-\alpha )(3-\beta) }(R,1/\lambda)\,, \\
 X^C_{\alpha \beta} (R,\lambda) &= X^C_{\beta \alpha }(R,\lambda) = X^C_{(3-\alpha )(3-\beta) }(R,1/\lambda)\,,\\
 Y^C_{\alpha \beta} (R,\lambda) &= Y^C_{\beta \alpha }(R,\lambda) = Y^C_{(3-\alpha )(3-\beta) }(R,1/\lambda)\,.
\end{align}
\end{subequations}
\citet{jeffrey1984calculation} provided series expansions in $1/R$  for the resistance functions using the twin-multipole method, as well as 
small-$s$ asymptotes. Following Ref.~\cite{dhanasekaran2021collision}, the two asymptotes are matched together using exponential smoothing as described in Section~\ref{sec:electrostatic} for the electrostatic force.

Townsend \cite{townsend2018generating} corrected a number of errors in the small-$s$ expressions derived by Jeffrey \& Onishi \cite{jeffrey1984calculation}. 
We used Townsend's corrections for the small-$s$ asymptotics of the functions $X_{12}^A, Y_{12}^B,  Y_{11}^C$ and $ Y_{12}^C$. In particular, the corrections mentioned by \citet{townsend2018generating} were used for the functions $A_{12}^X$, $B_{12}^Y$, $C_{11}^Y$, $C_{12}^Y$ and the function $g_5$ 
(see~\citet{townsend2018generating} for definitions of these functions, Eqs.~(15), (23), (30), (31), and (32) in that paper.
In order to verify that these corrections provide consistent approximations for the hydrodynamic forces and torques, we made the following checks. First, for all ten functions, we summed the large-$R$ series expansions derived by~\citet{jeffrey1984calculation}, 
 including $150$, $200$, $250 $,and $300$ terms, and confirmed that the results were consistent with the small-$s$ asymptotes.
 Second, we compared with the results for the radial and tangential mobility functions of Ref.~\cite{dhanasekaran2021collision}. 
We inverted Eq.~\eqref{eq:resistance}, and set hydrodynamic forces and torques to zero, to obtain the mobility functions for ${\rm St}\!=\!0$-droplets. We obtained the functions $L$ and $M$ shown in Figs. 5 and 9 of Ref.~\cite{dhanasekaran2021collision}. This constrains the resistance functions $X^A_{\alpha\beta}$, and a linear combination of $Y^A_{\alpha\beta}$ and $Y^B_{\alpha\beta}$.

When the interfacial separations between two droplets are comparable to the mean free path $\ell$ of air, the hydrodynamic approximation breaks down. Instead,
 the droplets move in a non-continuum flow described by the Boltzmann equation.  The resulting non-continuum corrections to the radial hydrodynamic forces between two spheres were first computed by \citet{sundararajakumar1996noncontinuum}. They used solutions to the linearised Boltzmann equation to calculate the non-continuum radial resistance functions for interfacial separations smaller than the mean free path. The corresponding tangential corrections were evaluated by \citet{li2021non}. In our  numerical computations we used the uniformly
valid resistance functions (denoted by the superscript ${\rm uv}$) quoted in Ref.~\cite{li2021numerical}, their Eqs.~(4.12) to (4.15):
\begin{subequations}
 \begin{align}
  X_{\alpha \beta}^{A,{\rm uv}} &= X_{\alpha \beta}^A +(-1)^{\alpha +\beta}a_{\alpha  \beta}\Big(2 \frac{f^{||}_{\rm fit}}{\rm Kn}-\frac{1}{\xi}\Big)\,,\\
  Y_{\alpha \beta}^{A,{\rm uv}} &= Y_{\alpha \beta}^A +(-1)^{\alpha +\beta}a_{\alpha  \beta} \Big[\frac{1}{3\sqrt{\pi}}\widetilde{W} +\Big(\frac{1-\lambda}{1+\lambda}\Big)^2 \widetilde Q\Big]\,,\\
  Y_{\alpha \beta}^{B,{\rm uv}} &= Y_{\alpha \beta}^B +2a_{\alpha  \beta} \Big( \frac{a_{\alpha}}{\overline{a}}\Big)^{|\alpha -\beta|} \Big[\frac{(-1)^\beta}{4\sqrt{\pi}}\widetilde{W} -(-1)^{(\alpha+\beta)}\frac{3}{4}\Big(\frac{1-\lambda}{1+\lambda}\Big)^2 \widetilde Q\Big]\,,\\
  Y_{\alpha \beta}^{C,{\rm uv}} &= Y_{\alpha \beta}^C +a_{\alpha  \beta} \Big( \frac{2 a_H}{\overline{a}}\Big)^{|\alpha -\beta|} \Big[\frac{1}{4\sqrt{\pi}}\widetilde{W} +(-1)^{(\alpha+\beta)}\frac{3}{4} \widetilde Q\Big]\,.
 \end{align}
\end{subequations}
The above equations describe corrections to 16 out of the 20 resistance functions mentioned above.The  radial rotational resistance function $X^C_{\alpha\beta}$ is not modified \cite{li2021numerical}.
The functions  $f^{||}_{\rm fit}, \widetilde W, \widetilde Q$ are given in Ref.~\cite{li2021numerical}. In the above equations, $\overline{a}$ is the mean droplet radius defined above, 
$a_H=a_1a_2/(a_1+a_2)$ is the harmonic mean, and $a_{\alpha\beta}=2a_H/(a_\alpha+a_\beta)$  \cite{li2021numerical}. The non-continuum resistance functions mentioned above are computed using the linearised Boltzmann equation, valid when the Mach number remains small. An additional simplification used is that the collision term in the Boltzmann equation is replaced by a collision operator which relaxes the molecular velocity to a Maxwellian distribution \cite{bhatnagar1954a}. Finally, the results are valid for asymptotically small Kn numbers.

The resistance functions coupling translations and rotations in Eq.~\eqref{eq:resistance} are subleading in $1/R$ to the translational resistance functions \cite{kim2005microhydrodynamics}. When the droplets are far apart, the angular velocities may be therefore neglected, as in Ref.~\cite{magnusson2021collisions}.
However,  for droplets traveling at small interfacial separations $s$ the angular dynamics must be considered, because hydrodynamic torques tend to cause one droplet to roll over the other one.

\subsection{Numerical integration}
\label{sec:num_int}
To make sure that droplets do not overlap in the numerical simulations when they travel at very small interfacial distances for finite times,
we followed 
Ref.~\cite{li2021numerical} and employed a  $4$-th order Runge-Kutta algorithm with an adaptive time step. Our prescription for the adaptive time
step differs slightly from the one  in Ref.~\cite{li2021numerical}. We used
$ \Delta t = \varepsilon \min\left\{{s}/{ V_R}, {s}/{V_T}\right\}$.
Here, $V_R$ and $V_T$ are the radial and tangential speeds, and $\varepsilon\ll1 $ is a 
numerical constant. The numerical results shown in Fig.~\ref{fig:fig2} were obtained for  $\varepsilon = 0.08$. We verified that reducing the numerical constant to $\varepsilon=0.001$ 
did
not change the results shown in Fig.~\ref{fig:fig2}({\bf d}). 

We recorded a collision and stopped the numerical integration as soon as the interfacial distance became smaller than  $10^{-14}$ m.  We verified that using a larger cutoff, $10^{-13}$,  did not change the results shown in Fig.~\ref{fig:fig2}({\bf d}). 
\section{Results}

\begin{figure}
\begin{overpic}[scale=0.55]{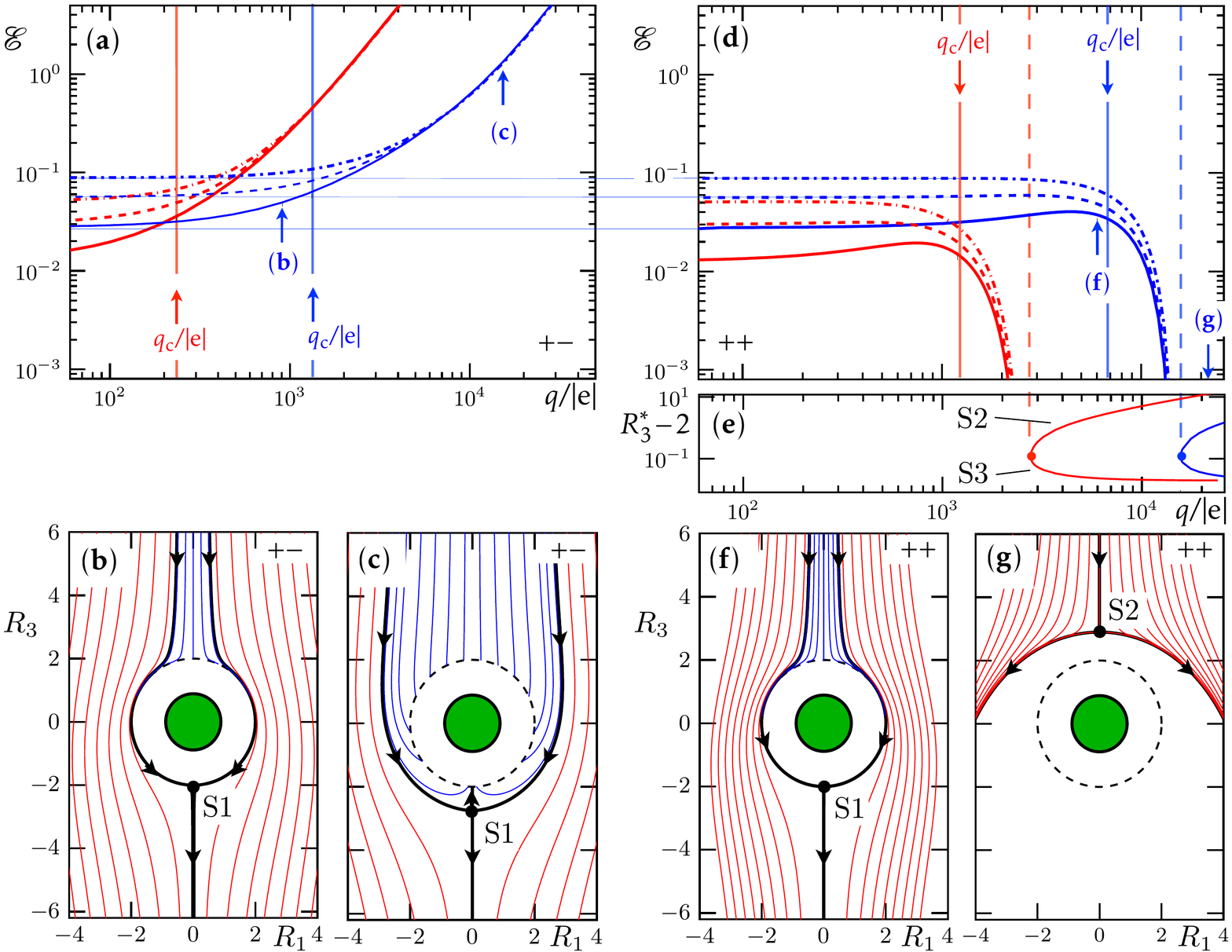}
\end{overpic}
\caption{\label{fig:fig2} Collision efficiencies and phase portraits of charged droplets settling in still air.
({\bf a}) Collision efficiency $\mathcal E$ for droplets with equal amounts of charge of opposite parity, as a function of charge  $q=|q_{1,2}|$ over elementary charge $e$. 
Droplet radii: $a_1 = 16 \,\mu$m and $a_2 = 20 \,\mu$m (blue), and $a_1 = 8 \,\mu$m and $a_2 = 10 \,\mu$m (red). 
Kn $=10^{-3}$ (solid lines), $10^{-2}$ (dashed), and $5\times10^{-2}$ (dash-dotted). The vertical solid lines  indicate the locations where Eqs.~\eqref{eq:theory_dim} and \eqref{eq:details} predict a transition. 
Arrows denote the charges in panels ({\bf b},{\bf c}).  Panel {(\bf b)} shows the relative droplet dynamics in the $R_1$-$R_3$-plane (where $R_i$ is non-dimensionalised by $\bar a$), for $a_1 = 16\,\mu$m, $a_2=20\,\mu$m, $q_{1,2}=\pm 908 \,e$, and Kn $=10^{-3}$. Colliding trajectories (blue), non-colliding trajectories (red), The solid black line
is the separatrix between colliding and non-colliding trajectories, and $\bullet$ represents the saddle point S$1$. ({\bf c}) Same as ({\bf b}), but for $q_{1,2}=\pm 15469\,e$. 
Manifolds of the saddle point S$1$ ($\bullet$) are shown as black solid lines.
({\bf d})~Same as ({\bf a}), but for equal parity. Panel ({\bf e}) shows the bifurcations (their locations are denoted by vertical dashed lines) that give rise to saddle points S$2$ and S$3$.  Panel ({\bf f}) shows the relative dynamics in the  $R_1$-$R_3$-plane for $a_1 = 16\,\mu$m, $a_2=20\,\mu$m, $q_{1,2}=6011\,e$ along with the saddle point S$1$, and panel ({\bf g}) corresponds to $q_{1,2}= 21199 \,e$, and shows the saddle point S$2$. 
} 
\end{figure}

In the following we discuss how the collision efficiency depends on the excess droplet charge.
The collision efficiency $\mathscr{E}$ is defined in terms of the critical impact parameter $b_{\rm c}$ beyond which the droplets cease to collide,
$\mathscr{E} = b_{\rm c}^2/(a_1+a_2)^2$.
 For droplets charged with equal and with opposite polarities,  we describe
 a transition from a regime at small charges that is dominated by non-continuum effects, to a charge-dominated regime at large excess charges. In both cases, the collision efficiency is determined by a saddle point
 below the collision sphere, and the transition between the regimes occurs when this saddle point leaves the region where non-continuum effects dominate [Fig.~\ref{fig:schematic}({\bf b})].
 Our results for the collision efficiency are shown in Fig.~\ref{fig:fig2} for droplets that carry the same amount of charge with different polarities (left), or with the same polarity (right). We discuss these two cases separately in  Sections \ref{sec:pm} and \ref{sec:samecharge}.

 The droplets were initialised 
sufficiently far apart that they settled independently at their steady Stokes 
settling velocities.  The initial angular velocities were set to zero. Since the settling velocities and the centre-of-mass separation $\ve R$
were initially in the $R_1$-$R_3$-plane, 
and since the torques and forces depend upon the cross product of the translational and angular velocities, respectively, with the separation vector \cite{kim2005microhydrodynamics},  the spatial dynamics of the droplets must remain confined to the $R_1-R_3$ plane. The only non-zero component of the angular velocities was $\omega_2^{(i)}$.  Thus, due to the choice of initial conditions, one component of the droplet velocities and two components of angular velocities remain zero, $v_2^{(i)} =0,\, \omega_1^{(i)} =0,$ and $\omega_3^{(i)} =0$.
 
\subsection{Opposite charges}
\label{sec:pm}
Figure~\ref{fig:fig2}({\bf a}) shows that electrostatic attraction increases the collision efficiencies $\mathscr{E}$ for droplet pairs of radii $8\,\mu$m and $10\,\mu$m (red) more strongly than for droplets with radii $16\,\mu$m and $20\,\mu$m (blue). The droplets were charged with the same amount of excess charge, but with different polarities.
For weak charges, $\mathscr{E}$ is approximately independent of charge, but the collision efficiency depends on the Knudsen number, Kn.
In this limit, the collision dynamics resembles that of neutral droplets settling in still air, as seen in Fig.~2({\bf a}) in Ref.~\cite{magnusson2021collisions} (but note that there are subtle differences that we discuss below).
For strong charges, by contrast, the collision efficiency is independent of Kn, as predicted by Magnusson {\em et al.} \cite{magnusson2021collisions}.
The difference between these two regimes can be understood by following the collision dynamics in the $R_1$-$R_3$-plane [Fig.~\ref{fig:schematic}({\bf a})],
corresponding to a frame that translates with the smaller droplet.

Consider first the limit of large charges, panel ({\bf c}). 
The collision efficiency is determined by the stable manifold of a saddle point S$1$ located below
the collision sphere. 
The unstable manifold of the saddle is one-dimensional, with a component along the $R_3-$axis. 
The stable manifold therefore has co-dimension one.  As a consequence, the stable manifold forms a separatrix for the phase-space dynamics  \cite{wiggins2013chaotic}. The corresponding solid black line in  panel ({\bf c}) represents the projection of a one-dimensional curve within the stable manifold, the curve defined by the chosen initial condition. Trajectories from above that approach the collision sphere above this line must collide (blue). All other trajectories from above do not collide (red). The saddle point  S$1$ has four complex eigenvalues. The corresponding eigenvectors span the $R_1-R_2-V_1-V_2$ space. This means that trajectories approaching the fixed point in this plane spiral around the fixed point in phase space. The trajectories plotted in black in panel ({\bf c}) do slightly overshoot the saddle point S$1$ , but this is not visible in the Figure.
Since the saddle point and its stable manifold are far from the collision sphere,  the local breakdown of the hydrodynamic approximation does not matter. 
As a consequence, the collision efficiency is independent of Kn. 
This is the limit analysed by Magnusson {\em et al.} \cite{magnusson2021collisions}, see their Fig.~2({\bf b}). 

The limit of small charges, shown in panel ({\bf b}), is more subtle than the large-charge limit. 
As mentioned above, the collision dynamics looks like that for neutral droplets, where the separatrices delineating collisions from no collisions  (solid black lines) are
grazing trajectories.  However, our numerics indicates that separatrices shown in panel ({\bf b}) do not graze the collision sphere. They lie very close to the collision sphere, but they  appear to connect to a saddle point S$1$ below the collision sphere, located on the $R_3$-axis. The numerical integration becomes difficult when the trajectory travels along the collision sphere at interfacial separations $s< {\rm Kn}$. Therefore we could not resolve the saddle point in the numerical integration (the trajectories corresponding to the black solid lines do not hit the saddle point, but they pass very close to it). 
As outlined above, a transition between the two different regimes shown in Fig.~\ref{fig:schematic}({\bf b}) occurs when the saddle point S$1$ exits the region $s< {\rm Kn}$. In order to compute the charge at which the transition occurs, we first start by determining the location of the saddle points of the relative dynamics. In order to find the saddle points of the relative droplet dynamics, consider the dynamical system defined by $\dot{\ve R}=\ve  v^{(2)}-\ve  v^{(1)}$, Eqs.~\eqref{eq:vdot} and \eqref{eq:model_c}. We look for zeros of this dynamical system, $\dot{\ve R}=0, \dot{\ve v}^{(i)}=0$ and $\dot{\ve \omega}^{(i)}=0$, $i=1,2$. At the saddle point, the droplets fall at a constant interfacial separation with a common steady-state settling velocity ${\ve v}^{(1)}={\ve v}^{(2)} = \ve v_{\rm s}$. Due to the symmetry of the problem, $\ve v_{\rm s}$ must point along the direction of gravity. Since there is no external torque acting on the droplets, and because of the dissipative nature of the dynamics, the angular velocities at the fixed point must vanish, ${\ve \omega}^{(i)}=0$. Thus, we find the location of this equilibrium point by requiring that \cite{magnusson2021collisions}
\begin{equation}
\label{eq:equ}
 \ve F^{(i)}_{\rm g}+\ve F^{(i)}_{\rm h} + \ve F^{(i)}_{\rm e} =0\quad \mbox{for}\quad i=1,2\,.
\end{equation}
Given the forces described in Section \ref{sec:model}, Eq.~(\ref{eq:equ}) simplifies to
\begin{subequations} 
\label{eq:equ2}
\begin{align}
 -m_1 g  - 6 \pi \mu \left( a_1 X^A_{11} +\overline{a}X^A_{12} \right) v_{\rm s} + {F}_{\rm e} &=0\,,\label{eq:saddle_a_appendix}\\
 -m_2g - 6 \pi \mu \left( a_2 X^A_{22} + \overline{a}X^A_{21} \right) v_{\rm s} -{F}_{\rm e} &=0\,.\label{eq:saddle_b_appendix}
\end{align}
\end{subequations}
Here $\mu = \varrho_{\rm f}\nu$ is the dynamic viscosity of air, and $v_{\rm s} \equiv |\ve{v}_{\rm s}|$ is the common settling speed of the two droplets in the steady state, and  $F_{\rm e} = F_{\rm e}^{(1)}$.
For given droplet radii and charges, solutions to these equations give the saddle-point location $[0,R_3^\ast]$, and centre-of-mass settling speed $V_{\rm s}$.
For the parameters corresponding to panel ({\bf b}), we find $s^\ast\approx 6\times 10^{-4}$ . So the equilibrium point is very close to, but not on the collision sphere. 
As mentioned above, we could not find the stable and unstable manifolds by numerical integration, because numerical integration is hard when the trajectories travel at interfacial separations $s<{\rm Kn}$ parallel to the collision sphere. However, the dynamics in the vicinity of the equilibrium point [panel ({\bf b})] indicates that it is a saddle: the fixed point has eigenvalues with positive and negative real parts. As the charge tends to zero, the saddle approaches the collision sphere, $R_3^\ast\to -2$.  Conversely, as the charge increases, 
the saddle moves down the $R_3$-axis.
We conclude that the saddle
point identified in Ref.~\cite{magnusson2021collisions} appears already at small Coulomb numbers, ${\rm Cu} \ll 1$. 
Since the numerical integration is more difficult at smaller (but non-zero) charges, we did not investigate this limit by numerical integration.

How much charge is needed to cross over from the small charge limit where non-continuum effects determine collisions to the charge-dominated limit? There  is no bifurcation,
the same saddle determines the collision dynamics in both regimes. But the above considerations indicate that the transition occurs when the saddle point leaves the region $s < {\rm Kn}$ in the $R_1$-$R_3$-plane,  the hashed region in Fig.~\ref{fig:schematic}({\bf b}). If the saddle point is close to the collision sphere, its location and  its manifolds depend on Kn. By contrast, when the interfacial separation at the saddle point is much larger than the mean free path $\ell$, the saddle point and its associated separatrix are independent of Kn, and the collision efficiency is determined by electrostatic forces. 

To find the critical charge, we set  the interfacial separation equal to the mean free path $\ell$ in Eqs.~(\ref{eq:equ2}), or
$s={\rm Kn}$ in non-dimensional units. Substituting $s={\rm Kn}$ into Eqs.~(\ref{eq:equ2}) and solving  for $V_{\rm s}$ and $F_{\rm e}$ we find the electrostatic force required to maintain a time-independent relative settling velocity $v_{\rm s}$,
\begin{subequations} 
\begin{align}
v_{\rm s}&=\frac{4 a_1^2 g \varrho_{\rm w} (1+\lambda^3)}{9 \mu \lambda^2}\frac{1}{2 \lambda X^{A,{\rm uv}}_{11} +(1+\lambda )(X^{A,{\rm uv}}_{12}+X^{A,{\rm uv}}_{21})+2X^{A,{\rm uv}}_{22}}\,,\\
 F_{\rm e}& = \frac{4 a_1^3 g \pi \varrho_{\rm w}}{3 \lambda^3}\frac{-2 \lambda X^{A,{\rm uv}}_{11} -(1+\lambda )X^{A,{\rm uv}}_{12}+\lambda^3 [(1+\lambda)X^{A,{\rm uv}}_{21}+2X^{A,{\rm uv}}_{22}]}{2 \lambda X^{A,{\rm uv}}_{11} +(1+\lambda )(X^{A,{\rm uv}}_{12}+X^{A,{\rm uv}}_{21}{)}+2X^{A,{\rm uv}}_{22}}\,. \label{eq:fe_c}
\end{align}
\end{subequations} 
To determine the corresponding amount of charge, we approximate  $F_{\rm e}$ and the resistance functions by their small-$s$ asymptotes (Section \ref{sec:model}).
We find
\begin{align} \label{eq:theory_dim}
 |q_1q_2| = \frac{1}{h(\lambda,{\lambda_q})}\frac{a_1^5 g \, {\rm Kn}  \rho_p}{k_e \lambda} \left[f^{(1)}(\lambda) \log \left( \frac{4\lambda}{(1+\lambda)^2 {\rm Kn}} \right)^2 + f^{(2)}(\lambda) \log \left( \frac{4\lambda}{(1+\lambda)^2 {\rm Kn}} \right)+ f^{(3)}(\lambda) \right].
\end{align}
In this expression, the function $h(\lambda,\lambda_q)$ is defined as
\begin{subequations}
\label{eq:details}
\begin{align}
 h(\lambda,{\lambda_q})&= \frac{1}{{|\lambda_q}|}\left[ \left(\psi(\tfrac{1}{1+\lambda})+\gamma\right) -\left(\psi(\tfrac{\lambda}{1+\lambda})+\gamma\right) {\lambda_q} \right]^2.
\end{align}
The  expressions for the functions $f^{(k)}$ are quite lengthy. Here we only quote  their Pad\'e approximants around $\lambda=1$, 
 of order $[3,3]$ (the transition values shown in Fig.~2 were obtained using the exact functions),
 \begin{align}
 f^{(1)} &= \frac{8.98831 (\lambda-1) - 12.553 (\lambda-1)^2 + 
 4.18158 (\lambda-1)^3}{1- 3.39659 (\lambda-1) + 3.95407 (\lambda-1)^2 -
 1.59769 (\lambda-1)^3}\,,\\
 f^{(2)} &= \frac{-45.6737 (\lambda -1)+2.16728 (\lambda -1)^2-0.229416 (\lambda -1)^3}{1+1.95255 (\lambda -1)+0.830927 (\lambda -1)^2-0.112841
   (\lambda -1)^3}\,,\\
  f^{(3)} &= \frac{-58.0221 (\lambda -1)-5.79319 (\lambda -1)^2-2.67774 (\lambda -1)^3}{1+2.09984 (\lambda -1)+1.39178 (\lambda -1)^2+0.352657
   (\lambda -1)^3}\,.
\end{align}
\end{subequations}
Fig.~\ref{fig:fig2} shows the predicted charge $q= |q_{1,2}|$ (assuming $\lambda_q = \pm 1$) computed using Eqs.~(\ref{eq:theory_dim}) and (\ref{eq:details}) as vertical  solid lines. We see that the theory predicts the transition from the small-charge limit where non-continuum effects determine collisions to the charge-dominated regime for both droplet sizes considered. This confirms the theoretical prediction that the transition occurs when the saddle point leaves the Kn-dominated region in the $R_1$-$R_3$-plane.

\subsection{Same charges}
\label{sec:samecharge}
Panel ({\bf d}) illustrates that the collision efficiencies $\mathscr{E}$ for droplet pairs with equal amounts of excess positive charge tend to zero as the charge
increases. Coulomb repulsion
 prevents collisions entirely at sufficiently large charges.
   Furthermore,  the collision efficiency vanishes more quickly for the droplet pair with radii $8\,\mu$m and $10\,\mu$m than for the droplet pair with radii $16\,\mu$m and $20\,\mu$m, as expected, because the Coulomb number is larger for smaller droplets with the same charge.
As in panel ({\bf a}), we identify two regimes. For small charges, the collision efficiency depends on Kn but the Kn-dependence becomes much weaker as the charge magnitude increases and the collision efficiency becomes charge dominated. The transition mechanism  is the same as for oppositely charged droplets: despite the fact that the droplets carry charges with the same polarity, the electrostatic force is attractive at small separations \cite{lekner2012electrostatics}. This gives rise to a saddle-point S$1$ below the collision sphere, as in panel ({\bf b}), where the attractive force balances differential settling. The corresponding predictions of Eq.~\eqref{eq:theory_dim} are shown as vertical solid lines in panel  ({\bf d}). 
One difference  to the collision dynamics for charged droplets with equal polarity [panel ({\bf a})] is that the crossover between the small-charge regime where non-continuum effects determine collisions  and the  charge-dominated regime occurs at much larger charges in panel ({\bf d}). The reason for this difference is that the small-$s$ attractive force between droplets with the same charge is weaker than the attractive force for droplets with opposite charges.  Moreover, the electrostatic force for droplets with the same charge is repulsive at large interfacial
distances. Since the force attracts at small separations, it must change sign at a non-dimensional critical distance  $s_{\rm c}$ that depends on $\lambda$ and $\lambda_q$.
Therefore, the  droplet interfacial separation at the saddle point must remain smaller than this critical distance, $s_{\rm c}$, as the charge increases.
For the parameters used here, $s_{\rm c} \approx 0.018$. This means
that the  collision efficiency for droplets with the same polarity becomes independent of the Kn number only when ${\rm Kn} < 0.018$. There is no such
condition for droplets charged with opposite charges, because the saddle point continues to move down as the amount of charge increases (Section \ref{sec:pm}).

As the  magnitude of the excess charge increases, a bifurcation  occurs above the collision sphere. This bifurcation gives rise to two new saddle points, S$2$ and S$3$. 
The saddle-point S$2$ results from the balance between electrostatic repulsion and differential settling. The stable manifold of S$2$ is $13$-dimensional. Its unstable manifold 
is two-dimensional,
therefore it does not form a separatrix for the collision dynamics. Nevertheless, the collision efficiency appears to vanish already before the bifurcation for reasons discussed below. In addition, droplets which approach the fixed point along the $R_3-$ axis spend a long time in each others vicinity.
The stable manifold of S$3$ is $12$-dimensional,  and its unstable manifold has dimension three. This saddle point is not approached by trajectories starting at large separations for the Stokes numbers considered here. However, at much larger Stokes numbers, trajectories could overshoot the saddle S$2$ and approach the fixed point S$3$.

We determined the charge $q_{\rm c}$ at which the bifurcation occurs by numerically solving  Eq.~(\ref{eq:equ2}) for the interfacial separation $s$ and the common settling 
speed $v_{\rm s}$. We determined the location of the bifurcation by computing the value of the droplet charge for which Eq.~\eqref{eq:equ2} had exactly one solution for $R_3 >0$. 
The locations of the two saddles on the $R_3$-axis, for $R_3 >0$,  are shown in panel ({\bf e}).
Panel ({\bf f}) shows the relative dynamics in the $R_1$-$R_3$-plane before bifurcation. It is qualitatively similar to panel ({\bf b}). 
 At weak charges, it does not matter whether the droplets have the same or opposite polarities. For both cases, a saddle point below results due to the attractive force at close approach.

Panel ({\bf g}) shows the trajectories in the $R_1$-$R_3$-plane after the bifurcation. 
The flow in the vicinity of S$2$ confirms that this fixed point is a saddle point. The second saddle point above, S$3$, and the saddle point  S$1$ below the collision sphere  are not shown, because these fixed points are not approached by trajectories approaching from afar. While the unstable manifolds
of the saddle point S$2$ do not form separatrices that prevent collisions from above, at the droplet sizes and initial conditions considered, we do not observe any collisions for charges above the bifurcation values. 
The bifurcation values for the charges are shown in panel ({\bf d}) as dashed vertical lines. 
However, we also observe that $\mathscr{E}$ tends to become vanishingly small well before the bifurcation occurs. The reason is that even before the saddle point appears, the dynamics in the $R_3-$direction slows down in its vicinity. These slow trajectories are then deflected in the unstable $R_1$ direction so that only trajectories approaching infinitesimally close to the $R_3$ axis can collide. We speculate that the qualitative form of the collision rate before the bifurcation can be understood by studying the normal form of the bifurcation for our system.

\section{Discussion}
The amount of excess charge required to reach the Coulomb-dominated collision regime depends on the polarity, upon whether the two droplets have excess charges of the same sign, or not.
For droplets carrying the same charges, larger amounts of charge are required  to qualitatively change the collision dynamics, about $q\sim 10^4\,e$ for $20\,\mu$m-droplets. For droplets that are charged with opposite polarities, much smaller charges are required, about $10^3\,e$ for $20\,\mu$m-droplets. 
In this case, the critical charge is much closer to the estimate of  \citet{davis1965theoretical} than that of \citet{tinsley2014comments}. 
However, these earlier estimates did not consider how the hydrodynamic and non-continuum forces vary at small interfacial distances. While \citet{davis1965theoretical} included hydrodynamic forces valid up to $10^{-3} a_2$ (where $a_2$ is the radius of the larger droplet), he did not account for non-continuum effects. As a consequence the
results for the collision efficiency depend on an arbitrary cutoff assumed to define a collision \cite{dhanasekaran2021collision}. 

\citet{tinsley2014comments} state that at least $10^4\,e$ are required for electrical forces to significantly affect the collision efficiency of oppositely charged droplets with
radii of the order of $20$ microns, and that for similarly charged droplets this threshold is even higher.  Our results indicate, by contrast, that the threshold is much smaller, about
 $10^3\,e$ for oppositely charged droplets, and about  $10^4\,e$ for droplets with the same polarity. 
 These differences are not surprising,
 given that early calculations \cite{paluch1970theoretical,schlamp1976numerical,schlamp1979numerical,tinsley2014comments} used highly idealised models that neither account for how the hydrodynamic forces change as the droplets come close, nor how hydrodynamics breaks down below the mean free path, 
 causing their predictions to fail \cite{abbott1974experimental}.  
 
 Eq.~\eqref{eq:theory_dim} shows how the critical charge $q_c$  depends upon the parameters of the problem. We see that $|q_1q_2| \sim a_1^5$. This means that the critical charge $q_c$ decreases rapidly as the droplet size decreases. Keeping all parameters the same as in Fig.~\ref{fig:fig2}, but with $a_2 = 4\,\mu$m   and $a_2 = 5\,\mu$m and Kn $=0.001$, we see that a charge of only $30 \,e$ is enough to lead to a transition for oppositely charged droplets. Note, however, that spatial
 diffusion (not considered here) must  affect the relative dynamics of such small droplets.

The model includes non-continuum effects due to the breakdown of hydrodynamics at small interfacial distances for both radial and tangential resistance functions. Ref.~\cite{li2021numerical} demonstrated that at small Stokes numbers, the collision efficiencies match those obtained for a St = 0 model from Ref.~\cite{dhanasekaran2021collision}, which did not include non-continuum corrections to the tangential resistance functions. However, the non-continuum corrections to tangential resistance functions might become important at larger St numbers, or when the radius ratio  of the droplets deviates significantly from unity \cite{li2021non}. 

Our model has five non-dimensional parameters: the Stokes number St (particle inertia), the Coulomb number Cu (charge), the ratio $\lambda$ of droplet radii, the charge ratio $\lambda_q$, and the Knudsen number Kn (effect of mean free path). We have not yet systematically studied the effect of changing St, $\lambda$, and $\lambda_q$. Instead we chose
$\lambda=0.8$, $\lambda_q={\pm}1$, and two values of the Stokes number, ${\rm St}=0.42$, and $3.96$. Since the droplet-size distribution in clouds at the onset of collisional growth is narrow \cite{grabowski2013growth}, it is natural to consider radius ratios close to unity, as we did here.
As a next step, it is of interest to study the effect 
of the parameter $\lambda_q$. While the charge distribution
of cloud droplets is not known in general, 
extreme values of $\lambda_q$ can significantly enhance collision rates \cite{khain2004rain}. Therefore it is important to study how the collision efficiency changes as the charge ratio $\lambda_q$ varies.

Let us briefly comment on the limitations of the model.
First, we did not consider van-der-Waals forces, short-range attractive dipole forces that change the collision dynamics at short interfacial distances $s$, of the order of the London length $\lambda_L$   
\cite{zhang1991rate}.  \citet{rother2022gravitational} computed the collision rate of droplets settling in still air including van-der-Waals forces
and  Maxwell slip (the first-order correction to the non-continuum effects).
They found that van-der-Waals forces cause a significant increase in the collision rate for droplets smaller than $10\,\mu$m in radius,
but that the effect is small for larger droplets.  This is consistent with the findings of \citet{dhanasekaran2021collision}, who stated that van-der-Waals forces {do not matter, compared to non-continuum effects,
for water droplets with radii larger than $10\,\mu$m. 
In order to understand how van-der-Waals forces affect the collision dynamics of charged droplets, we included the van-der-Waals force in our simulations, using \cite{ho1968preferential,zhang1991rate}
 \begin{align}
 \label{eq:vdw}
 F_{\rm vdW}^{(i)}  = -{A_{\rm H}} \begin{cases}
         \frac{{a_1 a_2}}{a_1+a_2} \frac{\lambda_L(\lambda_L+7.0768\pi s)}{s^2(\lambda_L+2\pi (1.7692) s)^2} &\quad \mbox{for}\quad s < 0.25\,\lambda_L/\pi\,,\\
        \frac{{a_1 a_2}}{a_1+a_2} \left( \frac{4.9 \lambda_L}{60\pi s^3}-\frac{6.51 \lambda_L^2}{360\pi^2 s^4}+\frac{2.36 \lambda_L^3}{1680\pi^3 s^5}\right) &\quad  \mbox{for}\quad s \geq 0.25\,\lambda_L/\pi\,.\\ 
      \end{cases}
 \end{align}
 Here $A_H$ is the Hamaker constant, with typical value $A_H=5\times 10^{-20}$ J for water droplets in air. 
 Two new non-dimensional parameters come with Eq.~(\ref{eq:vdw}): $\lambda_L/{\overline a}$ quantifies the interfacial separation below which van-der-Waals forces become important, while $A_H/(\tilde{m} V_0^2)$ quantifies the strength of van-der-Waals forces with respect to the relative kinetic energy at large separations.

 We found that including van-der-Waals forces in our simulations 
leads to a small  increase of about 15\% in the collision efficiency for Kn$=0.01$ and droplets of sizes $16\,\mu$m and $20\,\mu$m, but does not cause qualitative changes to Figs.~\ref{fig:fig2}({\bf a,d}). This is consistent with 
the findings of Refs.~\cite{dhanasekaran2021collision,rother2022gravitational}.
  In the limit of small charges and small Kn, by contrast, van-der-Waals forces may change of the collision dynamics qualitatively. We expect 
van-der-Waals force gives  rise to  a saddle point below the collision sphere, even for neutral droplets. In this case, the stable manifold of this new fixed point may determine the collision efficiency at small values of Kn.

Second, our model uses the Stokes approximation for the hydrodynamic forces and torques. The effects of convective and unsteady fluid inertia were neglected. 
\citet{klett1973theoretical} explained that convective fluid inertia makes a significant difference to the  collision efficiency of droplets with radius ratios close to unity, because the droplet dynamics in the Stokes approximation is degenerate when $a_1=a_2$ (see also Ref.~\cite{candelier2016settling}).
\citet{magnusson2021collisions} considered the leading-order effects of the particle Reynolds number and Strouhal number for droplets with ${\rm Re}_p \approx 0.1$ and ${\rm Sl} \approx 0.1$, at large droplet separations. Here we do not consider these effects because  we do not know how to account for the combined effects of the Reynolds and Strouhal effects at small interfacial distances.

Third, we assumed that the droplets remain spherical before they collide. Droplet deformation during collisions is controlled by the Weber number We, defined as the ratio of the droplet kinetic energy upon contact  to its surface energy  \cite{qian1997regimes}. When the Weber number is much smaller than unity, as for a $20 \,\mu$m droplet settling at its terminal velocity (We $ \sim 10^{-6}$), droplet deformation is negligible. 
At much larger collision speeds 
the Weber number is larger. In this case, and the droplets are expected
to deform. This changes not only the electrostatic forces, but also the hydrodynamic forces and their non-continuum regularisations, effects not considered here.

Fourth, we assumed that the droplets are perfect conductors. The time scale  $\tau$ of  charge redistribution on a water droplet depends on the ratio of the permittivity of water $\varepsilon_{\rm w}$ to its electrical conductivity $\sigma$, $\tau = \varepsilon_{\rm w}/\sigma$ \cite{haus1989electromagnetic}.
With the conservative estimate $\sigma \sim 5 \times 10^{-6}$ S/m for pure water, and $\varepsilon_w \sim 7 \times 10^{-10}$ C$^2$/Nm \cite{crc103},
one finds
$\tau \sim 1.4 \times 10^{-4}$~s. 
This time is much shorter than  the particle-relaxation time for relative dynamics, $\tau_\gamma= |\gamma_2^{-1}-\gamma_1^{-1}|$.
For   a pair of $16\, \mu$m- and $20\,\mu$m-droplets, one finds
 $\tau/\tau_\gamma \sim 0.08$.  The small but non-zero charge-relaxation may still have a small effect on the electrostatic forces which we neglected.

Fifth, we considered special initial conditions. We assumed that the droplets are initially so far apart that they settle independently with
their respective Stokes settling speeds, and we set their initial angular velocities to zero. As a result, the separatrices shown in Fig.~\ref{fig:fig2} correspond to  intersections between the invariant manifolds and the one-parameter family of curves determined by the initial conditions. 

Sixth, we considered droplets settling in still air. What is the effect of turbulence on the collision efficiency of charged droplets? 
\citet{saffman1956collision} explained how turbulent strains increase collision rates of cloud droplets, neglecting
their interactions. \citet{dhanasekaran2021collision} analysed how hydrodynamic interactions and their regularisation due to the breakdown of hydrodynamics affect
the collision dynamics of droplets settling in a steady straining flow. They found much smaller collision rates than Saffman \& Turner, and that hydrodynamic 
interactions lead to an intricate dependence of the collision rate on the straining flow. This behaviour is explained by a sequence of  bifurcations, both bifurcations of equilibria and grazing bifurcations \cite{dubey2022bifurcations}. It remains an open question how electrostatic interactions change this picture. Moreover, the close approach of charged droplets in turbulence was measured in Refs.~\cite{lu2010clustering_a,lu2010clustering_b,lu2015charged}, and compared  with a theory valid for large droplet separations -- similar to
the model described by \citet{magnusson2021collisions}. In order to quantitatively describe the relative dynamics of nearby droplets, we intend to generalise the model
to arbitrary linear flows. In the presence of a background flow, convective-inertia effects may give rise to inertial  lift forces  \cite{candelier2021time}.

\section{Conclusions}
We analysed the collision dynamics of weakly charged, $\mu$m-sized water droplets settling in still air by numerical integration of a model that incorporates particle inertia, hydrodynamic interactions in the Stokes approximation, non-continuum effects, and electrical forces.
We observed two distinct collision regimes. For small charges, the collision dynamics is dominated by short-range hydrodynamic interactions. In this regime it is crucial
to take into account how these interactions are  regularised below the mean free path of air.  As a consequence, the collision efficiency depends on the Knudsen number, Kn,
in this regime.
At large charges, in contrast, the collision efficiency does not depend on Kn, because the separatrix between colliding and non-colliding trajectories is the stable manifold of a saddle point far from the collision sphere, consistent with the conclusions drawn in Magnusson {\em et al.} \cite{magnusson2021collisions}.

The two regimes shown in Fig.~\ref{fig:schematic}({\bf b}) occur both for droplet pairs with excess charges of the same and of different polarities. This may be surprising at first sight,
considering that equal point charges repel each other, while opposite point charges attract. The reason is
that induced charges cause the electrostatic force to always be attractive, at small enough separations \cite{lekner2012electrostatics}. Ref.~\cite{khain2004rain} describes how induced electrical charges enhance the collision rate between a charged droplet and neutral droplet,
where the effect of induced charges is apparent at large separations 
(where their model applies). 
For micron-sized droplets of similar sizes, the electrostatic force changes sign at an interfacial distance much smaller than the droplet radius. But since this distance is of the order
of the mean free path, where the hydrodynamic approximation breaks down, the induced charges have a significant effect on the collision efficiency,
as we demonstrated in this paper.

We found that the saddle point S$1$ (Fig.~\ref{fig:fig2}) below the collision sphere exists even at small charges. But when the interfacial distance at the saddle-point $s < {\rm Kn}$, its stable manifolds depend on the Knudsen number. 
As a result the collision efficiency depends on Kn, at small charges. 
 The crossover between the two limits occurs when the saddle point moves further away from the collision sphere than the mean free path.  
 The critical charge where this qualitative
 change in the collision dynamics occurs is much lower than that stated by \citet{tinsley2014comments}. While our prediction matches  that of \citet{davis1965theoretical}, they did not include non-continuum effects at small separations, so that their result depends upon the cutoff used to define a collision. 
 We observed that for $20\,\mu$m-droplets, the transition between the two regimes happens at $\sim 10^3\,e$ for droplets charged with opposite polarities,  and at $\sim 10^4\,e$ for droplets charged with the same polarities. These charge magnitudes are comparable to those observed for thunder-cloud droplets \cite{takahashi1973measurement}. Note, however, that in thunder clouds, electric fields affect the droplet dynamics. Here we did not consider this effect.
   
 These conclusions are based on a model with five non-dimensional parameters, and we have only analysed a small part of the parameter space.  Including van-der-Waals forces give rise to two additional dimensionless parameters \cite{zhang1991rate,rother2022gravitational}, one quantifying the interfacial separation below which the van-der-Waals force is important, and another quantifying its strength.  While previous studies \cite{dhanasekaran2021collision,rother2022gravitational} suggest that including van-der-Waals force does not give rise to a qualitative changes in the dynamics for the parameters typical of cloud droplets with radii larger than $10\,\mu$m, we speculate that the qualitative dynamics are different when the Kn number becomes vanishingly small. The reason is that the attractive van-der-Waal force gives rise to a saddle point below, which determines the relative dynamics for vanishingly small Kn numbers. 

 Finally,  we assumed here that the droplets settle in quiescent air. We did not consider the effects of turbulent flow. Turbulence can substantially increase the rate at which droplets collide \cite{saffman1956collision,dhanasekaran2021collision}, because turbulent strains bring similar-sized droplets together. 
 More generally, turbulence increases spatial clustering of droplets, as measured by the pair-correlation function  (see Ref.~\cite{gustavsson2016statistical}
 for a review). Lu {\em et al.}  \cite{lu2010clustering_a,lu2010clustering_b,lu2015charged} measured how electrical charges change the pair-correlation function, and developed models to explain the observed effects -- at separations much larger than the droplet radii where hydrodynamic interactions can be neglected. It would be of interest to include in the model the mechanisms described here, in order to determine how charges affect the pair-correlation function at smaller separations. Similarly, the relative-velocity statistics of charged droplets in turbulence was investigated in Ref.~\cite{alipchenkov2004clustering}, neglecting hydrodynamic interactions, but our results indicate that these interactions (and their regularisation by non-continuum effects) cannot be neglected at small separations. In summary, the  phase-space picture presented here shows that  fairly small charges can have a significant effect on the relative dynamics of nearby droplets. How this affects the pair-correlation function, relative-velocity statistics, and collision efficiencies in turbulence remains an open question.

 {\em Acknowledgements}.
 The research of AD and BM was supported by grants from Vetenskapsr\aa{}det (grant nos.~2017-3865 and 2021-4452 and from the Knut and Alice Wallenberg Foundation (grant no.~2014.0048). The research of GB was supported by the National Science Foundation under grant no.~CBET-1605195. BM acknowledges support from a  Mary Shepard B. Upson Visiting Professorship with the Sibley School of Mechanical and Aerospace Engineering at Cornell.


%

\end{document}